\documentclass[conference]{IEEEtran}
\IEEEoverridecommandlockouts
\usepackage{cite}
\usepackage{amsmath,amssymb,amsfonts}
\usepackage{algorithmic}
\usepackage{graphicx}
\usepackage{textcomp}
\usepackage{xcolor}

\usepackage{xurl}
\usepackage{multirow,times,booktabs,tabularx, bm, array}
\usepackage[colorlinks,linkcolor=red]{hyperref}

\def\BibTeX{{\rm B\kern-.05em{\sc i\kern-.025em b}\kern-.08em
    T\kern-.1667em\lower.7ex\hbox{E}\kern-.125emX}}
\title{Graph-Enhanced Dual-Stream Feature Fusion with Pre-Trained Model for Acoustic Traffic Monitoring}

\begin{document}
\author{
    \IEEEauthorblockN{
        Shitong Fan$^{1\dagger}$,
        Feiyang Xiao$^{1\dagger}$,\thanks{$\dagger$These authors contributed equally to this work.}
        Wenbo Wang$^{2}$,
        Shuhan Qi$^{3}$,
        Qiaoxi Zhu$^{4}$, 
        Wenwu Wang$^{5}$,
        Jian Guan$^{1*}\thanks{*Corresponding author.}$}
    \IEEEauthorblockA{
        $^1$Group of Intelligent Signal Processing, Harbin Engineering University, Harbin, China\\
        $^2$Faculty of Computing, Harbin Institute of Technology, Harbin, China\\
        $^3$School of Computer Science and Technology, Harbin Institute of Technology, Shenzhen, China\\
        $^4$Acoustics Lab, University of Technology Sydney, Ultimo, Australia\\
        $^5$Centre for Vision, Speech and Signal Processing, University of Surrey, Guildford, UK\\
        }
}
\maketitle
\begin{abstract}

Microphone array techniques are widely used in sound source localization and smart city acoustic-based traffic monitoring, but these applications face significant challenges due to the scarcity of labeled real-world traffic audio data and the complexity and diversity of application scenarios. The DCASE Challenge's Task 10 focuses on using multi-channel audio signals to count vehicles (cars or commercial vehicles) and identify their directions (left-to-right or vice versa). In this paper, we propose a graph-enhanced dual-stream feature fusion network (GEDF-Net) for acoustic traffic monitoring, which simultaneously considers vehicle type and direction to improve detection. We propose a graph-enhanced dual-stream feature fusion strategy which consists of a vehicle type feature extraction (VTFE) branch, a vehicle direction feature extraction (VDFE) branch, and a frame-level feature fusion module to combine the type and direction feature for enhanced performance. A pre-trained model (PANNs) is used in the VTFE branch to mitigate data scarcity and enhance the type features, followed by a graph attention mechanism to exploit temporal relationships and highlight important audio events within these features. The frame-level fusion of direction and type features enables fine-grained feature representation, resulting in better detection performance. Experiments demonstrate the effectiveness of our proposed method. GEDF-Net is our submission that achieved 1st place in the DCASE 2024 Challenge Task 10.

\end{abstract}
\section{Introduction}
\label{sec:intro}

Acoustic-based traffic monitoring uses roadway sounds to estimate vehicle counts, speeds, and types, aiding in traffic control and anomaly detection \cite{5djukanovic2020robust, 6djukanovic2021neural, can_damiano_2024,4bulatovic2022mel}. Typically, single sensors are deployed along the roads to capture audio data for estimating speeds \cite{4bulatovic2022mel} or counting vehicles \cite{5djukanovic2020robust,6djukanovic2021neural,can_damiano_2024}. While being effective for speed estimation \cite{ 4bulatovic2022mel}, single-sensor setups struggle with vehicle direction detection due to limited spatial resolution and challenges in distinguishing overlapping signals from multiple vehicles, which restricts the system’s ability for traffic flow monitoring. To address these limitations, microphone arrays are used to capture multi-channel signals from various positions, enabling direction detection \cite{1ishida2018saved, 8severdaks2013vehicle, 3szwoch2021acoustic, 7zu2017vehicle}. Configurations include equidistant \cite{8severdaks2013vehicle}, orthogonal \cite{3szwoch2021acoustic}, and circular arrays \cite{7zu2017vehicle}. However, their effectiveness is often compromised by the scarcity of labeled real-world traffic data.

\begin{figure*}[!ht]
  \centering
  \includegraphics[width=.92\textwidth]{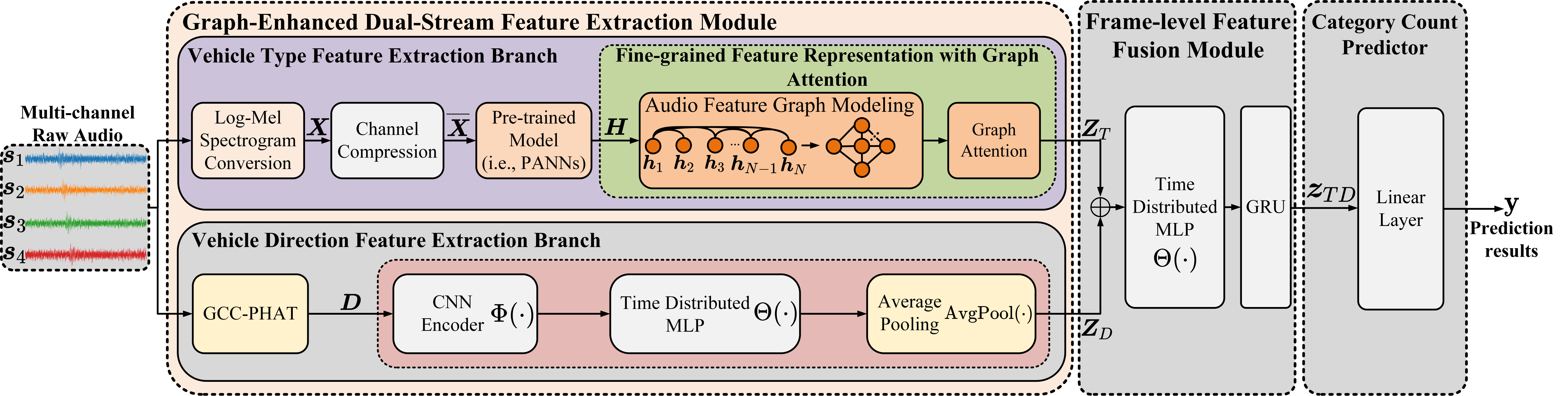}
  \vspace{-2mm}
  \caption{Overall framework of the proposed method. The proposed GEDF-Net includes a graph-enhanced dual-stream feature extraction (GEDF) module with a vehicle type feature extraction (VTFE) branch and a vehicle direction feature extraction (VDFE) branch, a  module for fusing type and direction features over time frames, and a category count predictor for vehicle prediction.}
  \label{fig:1}
  \vspace{-6mm}
\end{figure*}

To address these challenges, the Detection and Classification of Acoustic Scenes and Events (DCASE) 2024 Challenge introduces Task 10, Acoustic-based Traffic Monitoring \cite{can_damiano_2024,pyroadacoustics_damiano}, which focuses on detecting vehicle types (cars and commercial vehicles) and travel directions (left-to-right or vice versa). Due to the scarcity  of real-world data, the Challenge incorporates synthetic data to evaluate its impact on system performance \cite{pyroadacoustics_damiano}. The Task 10 baseline \cite{can_damiano_2024,Baseline2024dcaseT10} uses a dual-branch CNN-based network to extract vehicle direction and type features from Generalized Cross-Correlation with Phase Transform (GCC-PHAT) \cite{gccphat5670137} and log-Mel spectrograms, which are fused along the temporal dimension to detect vehicle events. To mitigate data scarcity, the baseline and other top-performing systems \cite{2Bai2024dcaseT10, 3Takahashi2024dcaseT10, 5Park2024dcaseT10, 6Betton2024dcaseT10, 7Cai2024dcaseT10}, including ours \cite{Guan2024dcaseT10}, all use pre-training on synthetic data \cite{pyroadacoustics_damiano} followed by fine-tuning on limited real data. The Top-3 system \cite{3Takahashi2024dcaseT10} improves the baseline by adding a matching loss \cite{relativehelmbold1999} to improve the alignment of the predictions with ground truth, achieving slightly enhanced performance.

Other systems, such as Top-4 \cite{5Park2024dcaseT10}, Top-5 \cite{6Betton2024dcaseT10}, and Top-6 \cite{7Cai2024dcaseT10}, utilize CNN-based architectures like ResNet \cite{resnethe2016deep} (Top-4) and VGG11 \cite{vggsimonyan2014very} (Top-6), with Top-5 also exploring ensemble methods. However, these methods do not outperform the baseline. A common limitation among these methods, including Top-3 \cite{3Takahashi2024dcaseT10}, is their limitation in capturing contextual relationships between audio events. In addition, while synthetic data supports feature learning, it does not fully resolve data scarcity and quality issues, limiting effective representation of the vehicle audio events.

The Top-2 system \cite{2Bai2024dcaseT10} combines vehicle direction (GCC) and type features (log-Mel spectrogram) along the temporal dimension and employs a Transformer \cite{transvaswani2017attention} for contextual modeling. However, directly using concatenated features without refinement can introduce redundancy and degrade performance. Like above methods \cite{can_damiano_2024, 3Takahashi2024dcaseT10, 5Park2024dcaseT10, 6Betton2024dcaseT10, 7Cai2024dcaseT10}, the Top-2 system \cite{2Bai2024dcaseT10} also adopts synthetic data to improve feature learning, which, however, can be limited by the quality of the data.

In this paper, we propose GEDF-Net, a graph-enhanced dual-stream feature fusion network with a pre-trained model, for acoustic-based traffic monitoring. The proposed GEDF-Net consists of a graph-enhanced dual-stream feature extraction (GDFE) module with a vehicle type feature extraction (VTFE) branch and a vehicle direction feature extraction (VDFE) branch for vehicle type feature and direction feature extraction, respectively, together with a frame-level fusion module to fuse the fine-grained features for enhanced vehicle detection. 

Specifically, in the VTFE branch, we use a pre-trained model (i.e. PANNs \cite{pannskong2020}, pretrained on AudioSet \cite{audiosetgemmeke2017}) to improve feature representation.  
In addition, inspired by \cite{graphac10098862}, we incorporate a graph attention mechanism \cite{graphcasanova2018} to capture temporal relationships between audio events, treating feature frames from PANNs as nodes and their relationships as edges, thereby enhancing vehicle type feature representation. The VDFE branch extracts direction features using GCC-PHAT for time delay estimation. Then, these features are integrated along the corresponding time dimension and fused at the frame level to obtain fine-grained representation for traffic monitoring. Finally, a category count predictor counts vehicles by type and direction (e.g., cars or commercial vehicles moving left-to-right or right-to-left). Experiments are conducted on the DCASE 2024 Challenge Task 10 dataset \cite{Baseline2024dcaseT10} to demonstrate the effectiveness of the proposed method, which achieved 1st place in DCASE Challenge.

\section{Proposed Method}
\label{sec:format}

Our proposed GEDF-Net, shown in Figure \ref{fig:1}, contains a dual-stream feature extraction module including two branches, namely, the VTFE branch for extracting vehicle type features and the VDFE branch for extracting direction features, together with a module to combine these features in frame-level for fine-grained representation and a category count predictor to estimate the number of vehicles in each category.

\subsection{Graph-Enhanced Dual-Stream Feature Extraction Module}

The GEDF module is used to extract the type and direction features, both from the four-channel audio, detailed as follows.

\subsubsection{Vehicle Type Feature Extraction Branch}

Inspired by \cite{graphac10098862}, the VTFE branch enhances the vehicle type feature representation by using a pre-trained model to address data scarcity and a graph attention mechanism to capture temporal relationships and emphasize important audio events for finer representation.

\noindent\textbf{Feature Enhancement with Pre-trained Model:} We use a pre-trained model (i.e., PANNs \cite{pannskong2020}) to extract vehicle type features. Since PANNs is trained on AudioSet \cite{audiosetgemmeke2017}, which includes vehicle data, it enhances vehicle type feature representation and helps address data scarcity by incorporating external knowledge.

To achieve this, we first convert the input four-channel raw audio signals $\bm{S} = \{\bm{s}_1, \bm{s}_2, \bm{s}_3, \bm{s}_4 \} \in \mathbb{R}^{4 \times L}$ to log-Mel spectrogram $\bm{X} \in \mathbb{R}^{4 \times B \times T}$ via a log-Mel spectrogram conversion operation, where $L$, $B$, and $T$ represent the number of sampling points, mel bins, and time frames, respectively.

Then, a convolution layer, i.e., $\text{Conv}(\cdot)$, is applied to compress the four-channel log-Mel spectrogram $\bm{X}$ into 1-D expression $\bm{\overline{X}} \in \mathbb{R}^{1 \times B \times T}$, as follows,
\begin{equation}
\label{eq:1}
\bm{\overline{X}} = \text{Conv}({\bm{X}}),
\end{equation}

After this, a pre-trained model (i.e., PANNs \cite{pannskong2020}) is utilized to  enhance the feature representation, mitigating the data scarcity for vehicle type feature extraction, as follows,
\begin{equation}
\label{eq:2}
\bm{H} = \text{PANNs}(\bm{\overline{X}}),
\end{equation}
where $\bm{H} = \{\bm{h}_1, \dots, \bm{h}_n, \dots, \bm{h}_N\} \in \mathbb{R}^{K \times N}$ denotes the enhanced feature, and $\bm{h}_n \in \mathbb{R}^{K \times 1}$ is the $n$-th feature frame. $K$ and $N$ are the dimension of the feature at each frame and the number of time frames, respectively.

\noindent\textbf{Fine-grained Feature Representation with Graph Attention:}
Since vehicle audio events usually span multiple time frames, and feature frames of the same vehicle type should have higher correlations. To capture these contextual associations, we use audio feature graph modeling with an attention mechanism to emphasize key audio events related to vehicle traveling, achieving finer vehicle type feature representation.

Let $\bm{h}_i$ and $\bm{h}_j$ be the feature frames (nodes) in $\bm{H}$. The correlation between these frames is represented by the weight $a_{ij}$ (attention coefficient) of the edge between $\bm{h}_i$ and $\bm{h}_j$, calculated by a learnable linear mapping, following \cite{graphac10098862} \cite{graphcasanova2018},
\begin{equation}
\label{eq:3}
a_{ij} = \text{Softmax}(\text{LeakyReLU}\left( \bm{e}^{\top} [\bm{M} \bm{h}_i \| \bm{M} \bm{h}_j] \right)),
\end{equation}
where $\bm{M} \in \mathbb{R}^{K\times K}$ is the learnable linear mapping, and $\bm{e} \in \mathbb{R}^{2K\times 1}$ is the learnable attention vector. $\|$ denotes concatenation operation. 

We can then obtain an adjacency graph $\mathcal{A} \in \mathbb{R}^{N \times N}$ from $\bm{H} $, with its element $a_{ij}$ at $i$-th row and $j$-th column to represent the relation between feature nodes $\bm{h}_i$ and $\bm{h}_j$. Then, by aggregating the feature nodes of $\mathcal{A}$, we can obtain the improved type feature representation $\bm{Z}_{T} \in \mathbb{R}^{K \times N}$ as follows,
\begin{equation}
\label{eq:4}
\bm{Z}_{T} = \mathcal{A} \bm{H} \bm{M}^{\top} + \bm{H},
\end{equation}

\subsubsection{Vehicle Direction Feature Extraction Branch}
We also adopt GCC-PHAT for direction feature extraction in the VDEF branch following \cite{can_damiano_2024}. In addition, an average pooling operation is introduced to further explore important directional information, facilitating the fusion of the vehicle type feature and direction feature over time dimension.

Specifically, the short-time Fourier transform (STFT) is employed to obtain the phase  $\bm{P}_c \in \mathbb{R}^{F \times T}$ and $\bm{P}_k \in \mathbb{R}^{F \times T}$ of audio signals for each channel pair $c$ and $k$, where $\{(c, k) \mid c, k \in \{1, 2, 3, 4\}, \text{and} \ c \neq k\}$. $F$ denotes the number of frequency bins. Then, the time delay of the audio signals for each channel pair can be calculated as follows,
\begin{equation}
\bm{D}_{c,k} = \mathcal{F}^{-1}(\exp(j \cdot \text{Angle}(\bm{P}_{c} \odot \bm{P}_{k}^*))),
\end{equation}
where $*$ denotes the conjugate operation, $\odot$ represents element-wise multiplication, and $\text{Angle}(\cdot)$ computes the phase angle. $\mathcal{F}^{-1}(\cdot)$ stands for the inverse Fourier transform. Here, $Q$ is the number of GCC-PHAT coefficients calculated from the two signals. Thus, the time delay estimation for all the four-channel audio signals can be represented as 
$\bm{D} = \{ \bm{D}_{c,k} \mid c, k \in \{1, 2, 3, 4\}, \text{and} \ c \neq k \}$.

Finally, the direction feature representation $\bm{Z}_{D}\!\in\!\mathbb{R}^{K \times N}$ can be obtained via a convolutional encoder (i.e., $\Phi(\cdot)$) and an MLP (i.e., $\Theta(\cdot)$) with the time delay estimation as,
\begin{equation}
\label{eq:8}
\bm{Z}_D =\text{AvgPool}(\Phi(\Theta(\bm{D}))),
\end{equation} 
where AvgPool($\cdot$) denotes the average pooling operation.

\subsection{Frame-level Feature Fusion Module}

After extracting the vehicle type and direction features, we use a module to combine these features over each time frame to obtain a fine-grained representation accounting for both vehicle type and direction as follows,
\begin{equation} 
\label{eq:9} 
    \bm{z}_{TD} = \text{GRU}(\Theta([\bm{Z}_{T} \parallel \bm{Z}_{D}])), 
\end{equation}
where GRU($\cdot$) denotes the gated recurrent unit (GRU) \cite{grucho2014properties}, and $\bm{z}_{TD} \in \mathbb{R}^{1 \times d} $ is the last time step result of GRU for regressing the vehicle count, and $d$ denotes the dimension of the result for each time frame.

\subsection{Category Count Predictor}
A linear layer is utilized as the category count predictor to estimate the counts for each category of vehicle event across vehicle type (i.e., car and commercial vehicle) and travel direction (i.e., left to right and right to left), as follows,
\begin{equation}
\label{eq:10}
    \mathbf{y}= \text{ReLU}(\bm{W} \bm{z}_{TD} + \bm{b}),
\end{equation}
where $\mathbf{y} \in \mathbb{R}^{1 \times 4}$ is the prediction result for acoustic traffic monitoring, and $\bm{W}$ and $\bm{b}$ are the weight matrix and bias of the predictor, respectively.

\begin{table}[htbp]
  \caption{Metadata summary for each location.}
  \label{tab:1}
  \vspace{-3mm}
  \centering
  \resizebox{0.46\textwidth}{!}{
  \begin{tabular}{l c c c c c c}
    \toprule
    \textbf{location} & \textbf{loc1} & \textbf{loc2} & \textbf{loc3} & \textbf{loc4} & \textbf{loc5} & \textbf{loc6}\\
    \midrule
    \textbf{number of audio samples}                   &1256   & 56    & 4129  & 17 & 96 & 1740 \\
    \textbf{max-pass-by-speed}      &100    & 50    & 50    & 50 & 40 & 90 \\
    \textbf{max-traffic-density (per minute)}    &1000   & 900   & 500   & 400 & 140 & 900 \\
    \textbf{number of vehicle audio events}    &11515 & 901 & 21685 & 63 & 165 & 19116 \\
    \bottomrule
  \end{tabular}}
  \vspace{-3mm}
\end{table}

\begin{table*}[htbp]
    \centering
    \caption{Effectiveness validation of our proposed GEDF-Net on DCASE 2024 Challenge Task 10 development dataset, where Kendall's Tau Rank Corr (Kendall), RMSE and Ranking Score are used for evaluation. Note that, the ranking score reflects results that are only based on comparisons among the methods listed in this table.}
    \label{tab:overall}
    \vspace{-3mm}
    \resizebox{0.99\textwidth}{!}{
    \begin{tabular}{@{}ccccccccccccccccccccccccc@{}}
        \toprule
        \multirow{2}{*}{Methods}  & \multirow{2}{*}{Category} & \multicolumn{2}{c}{loc1} & \multicolumn{2}{c}{loc2} & \multicolumn{2}{c}{loc3} & \multicolumn{2}{c}{loc4} & \multicolumn{2}{c}{loc5} & \multicolumn{2}{c}{loc6} & \multirow{2}{*}{Ranking Score}\\
        \cmidrule(r){3-4} \cmidrule(r){5-6} \cmidrule(r){7-8} \cmidrule(r){9-10} \cmidrule(r){11-12} \cmidrule(r){13-14} 
        & & Kendall & RMSE & Kendall & RMSE & Kendall & RMSE & Kendall & RMSE & Kendall & RMSE & Kendall & RMSE \\ 
        \midrule
        \multirow{4}{*}{Baseline}        & car\_left    &\textbf{0.445} & \textbf{2.555} & 0.579 & 3.074 & 0.543 & 1.731 & 0.195 & 1.997 & 0.575 & 0.693 & 0.804 & 1.628 & \multirow{4}{*}{2.71} \\
                             & car\_right &0.423 & 2.978 & 0.337 & 2.917 & 0.569 & 1.294 & 0.038 & 1.674 & 0.371 & 0.693 & 0.700 & 1.822& \\
                            & cv\_left   &0.084 & 0.918 & 0.044 & 0.813 & 0.034 & 0.309 & 0.000 & 0.655 & \textbf{0.068} & 0.362 & \textbf{0.763} & \textbf{0.509} \\
                            & cv\_right  & 0.076 & 0.882 & 0.051 & 0.604 & 0.322 & 0.212 & 0.000 & 0.463 & 0.257 & 0.252 & \textbf{0.641} & \textbf{0.530} \\
        \midrule
        \multirow{4}{*}{GEDF-Net (w/o -P)} & car\_left    & 0.410   & 2.654 & 0.630 & 2.510 & \textbf{0.555} & \textbf{1.716} & 0.049 & 2.295 & \textbf{0.582} & \textbf{0.629} & 0.805 &1.646 &\multirow{4}{*}{2.52}\\
                            & car\_right & 0.440 & 2.920 & \textbf{0.516} & \textbf{2.239} & 0.572 & 1.281 & -0.063 & 2.882 & 0.394 & \textbf{0.679} & \textbf{0.704} & \textbf{1.810} \\
                            & cv\_left   & 0.176 & 0.909 & -0.034 & 0.850 & \textbf{0.174} & \textbf{0.307} & 0.000 & 0.655 & 0.045 & \textbf{0.351} & 0.690 & 0.601 \\
                            & cv\_right  & 0.117 & 0.937 & -0.051 & 0.717 & 0.299 & 0.212 & 0.000 & 0.463 & \textbf{0.361} & 0.238 & 0.521 & 0.648 \\
        \midrule
        \multirow{4}{*}{GEDF-Net (w/o -G)} & car\_left    &0.393 & 2.765 & 0.700 & \textbf{2.157} & 0.548 & 1.724 & \textbf{0.634} & \textbf{1.174} & 0.525 & 0.745 & 0.811 & \textbf{1.549} &\multirow{4}{*}{2.58}\\
                            & car\_right & 0.441 & 2.970 & 0.474 & 2.544 & 0.575 & 1.289 & \textbf{0.341} & \textbf{0.958} & 0.397 & 0.694 & 0.694 & 1.837 \\
                            & cv\_left   &0.172 & 0.962 & 0.143 & 0.798 & 0.009 & 0.314 & \textbf{0.296} & 0.607 & -0.047 & 0.361 & 0.743 & 0.582 \\
                            & cv\_right  &\textbf{0.126} & 0.954 & 0.058 & 0.726 & -0.002 & 0.222 & 0.120 & 0.613 & -0.083 & 0.300 & 0.611 & 0.576 \\
        \midrule
        \multirow{4}{*}{GEDF-Net}    & car\_left    & 0.434	&2.600	&\textbf{0.719}	&2.177	&0.551	&1.729	&0.097	&2.095	&0.557	&0.708	&\textbf{0.816}	&1.582 & \multirow{4}{*}{\textbf{2.04}}\\
                            & car\_right & \textbf{0.448}	&\textbf{2.919}	&0.401	&2.666	&\textbf{0.577}	&\textbf{1.275}	&0.240	&1.548	&\textbf{0.401}	&0.697	&0.684	&1.910 \\
                            & cv\_left   & \textbf{0.207}	&\textbf{0.892}	&\textbf{0.226}	&\textbf{0.783}	&0.171	&0.315	&0.182	&\textbf{0.604}	&0.058&0.362	&0.683	&0.604 \\
                            & cv\_right  & \textbf{0.126}	&\textbf{0.861}	&\textbf{0.171}	&0.677	&\textbf{0.377}	&\textbf{0.195}	&\textbf{0.445}	&\textbf{0.428}	&0.357	&\textbf{0.208}	&0.570	&0.594 \\
        \bottomrule
    \end{tabular}
    }
    \vspace{-5mm}
\end{table*}

\begin{table}[th]
  \caption{Performance comparison with top-ranking systems on the DCASE 2024 Challenge Task 10 evaluation set.}
  \label{tab:2}
  \vspace{-3mm}
  \centering
  \resizebox{0.46\textwidth}{!}{
  \begin{tabular}{l c c}
    \toprule
    \textbf{Methods} & \textbf{Official Ranking} & \textbf{Ranking Score }\\
    \midrule
     GEDF-Net (Ours) \cite{Guan2024dcaseT10} &\textbf{1}& \textbf{3.98} \\
          Bai\_JLESS\_task10\_1 \cite{2Bai2024dcaseT10} &2& 4.44 \\
  Takahashi\_TMU-NEE\_task10\_1 \cite{3Takahashi2024dcaseT10} &3& 4.77 \\
       Baseline\_Bosch\_task10 \cite{Baseline2024dcaseT10} &-& 5.17 \\
            Park\_KT\_task10\_3 \cite{5Park2024dcaseT10} &4& 5.67 \\
Betton-Ployon\_ACSTB\_task10\_1 \cite{6Betton2024dcaseT10} &5& 7.89 \\
           Cai\_NCUT\_task10\_1 \cite{7Cai2024dcaseT10} &6& 8.14 \\
    \bottomrule
  \end{tabular}}
  \vspace{-1mm}
  \begin{flushleft}
    \footnotesize{~~~*Official ranking results of DCASE 2024 Challenge Task 10.}
  \end{flushleft}
  \vspace{-7mm}
\end{table}

\section{Experiments and Results}
\label{sec:pagestyle}
\subsection{Experimental Setup}

\subsubsection{Dataset}
Since the official evaluation set is not available, we evaluated performance on the DCASE 2024 Challenge Task 10 development dataset \cite{Baseline2024dcaseT10}, which includes four audio event types (“car\_left”, “car\_right”, “cv\_left”, “cv\_right”) from six locations (loc1 to loc6) and synthetic data generated by an acoustic traffic simulator \cite{can_damiano_2024,pyroadacoustics_damiano}. Real data were recorded using a linear microphone array parallel to traffic flow. Sample counts per location are in Table \ref{tab:1}. Similar to the baseline method, synthetic data are also included in our training process.

\subsubsection{Evaluation Metrics}

Following the official baseline \cite{can_damiano_2024}, we use Kendall's Tau Rank Correlation (Kendall's Tau Corr) and Root Mean Square Error (RMSE) as evaluation metrics. Kendall's Tau Corr measures the ordinal association between predictions and actual results, while RMSE quantifies prediction errors. We also use a third metric, Ranking Score, as defined in the official Task 10 evaluation.

The Ranking Score evaluates systems based on average rankings across $6\times4\times2$ comparisons: 6 locations (loc1 to loc6), 4 audio event types (“car\_left,” “car\_right,” “cv\_left,” “cv\_right”), and 2 metrics (Kendall's Tau Corr and RMSE). Each comparison is assigned a ranking score, with Rank 1 being the best and Rank N the worst. The final performance is the average of all rankings, where a lower score indicates better performance.

\subsection{Effectiveness Validation and Analysis}

We conduct ablation experiments on the development set to validate the effectiveness of our proposed GEDF-Net. We compare the full method with two variants: GEDF-Net without PANNs (w/o -P) and without graph attention (w/o -G), along with the baseline \cite{Baseline2024dcaseT10,can_damiano_2024}. Results are in Table \ref{tab:overall}.

All our methods outperform the baseline, with GEDF-Net achieving the best ranking score, demonstrating its effectiveness in using pre-trained models for feature enhancement and graph attention for detailed feature representation. GEDF-Net significantly outperforms GEDF-Net (w/o -P) in locations with very few samples (e.g., loc2 with 56 samples and loc4 with 17 samples), highlighting the value of external knowledge in mitigating data scarcity.

In loc5, the traffic scenario is relatively simple, as characterized by low maximum traffic density (140 vehicles per minute) and low maximum pass-by speed (40 km/h). With a limited number of samples, GEDF-Net (w/o -P) still achieves comparable results to GEDF-Net.

Meanwhile, GEDF-Net outperforms GEDF-Net (w/o -G), showing the benefits of using graph attention modeling to capture temporal relationships and highlight important audio events in vehicle type features. However, at loc4, GEDF-Net (w/o -G) performs better, likely due to the limited number of audio events (63) at this location, which may limit the graph attention model to learn robust representations.

\subsection{Performance Comparison with the Top Ranking Systems}

The evaluation set in DCASE 2024 Challenge Task 10 was not released to the public, and most top-ranking systems were not open-sourced either. For this reason, we could not reproduce their results for comparison. Instead, we present the official evaluation results published by DCASE Challenge organisers to showcase the advantages of our method. The official ranking scores of the top systems are shown in Table \ref{tab:2}, with detailed results available on the competition website\footnote[1]{\url{https://dcase.community/challenge2024/task-acoustic-based-traffic-monitoring-results}}.

From Table \ref{tab:2}, it can be seen that our proposed GEDF-Net as the submission system outperforms all other systems, which shows the superiority of our proposed method, demonstrating the effectiveness of using the pre-trained model to mitigate the data scarcity and the graph attention to exploit temporal relationships and highlight important audio events for acoustic traffic monitoring. Moreover, our method surpasses the Transformer-based Top-2 system (i.e., Bai\_JLESS\_task10\_1 \cite{2Bai2024dcaseT10}) that simply uses log-Mel spectrogram as type feature, further illustrating the effectiveness of graph-enhanced fine-grained feature representation with the pre-trained model.

\begin{figure}[t]
  \centering
  \includegraphics[scale=0.25]{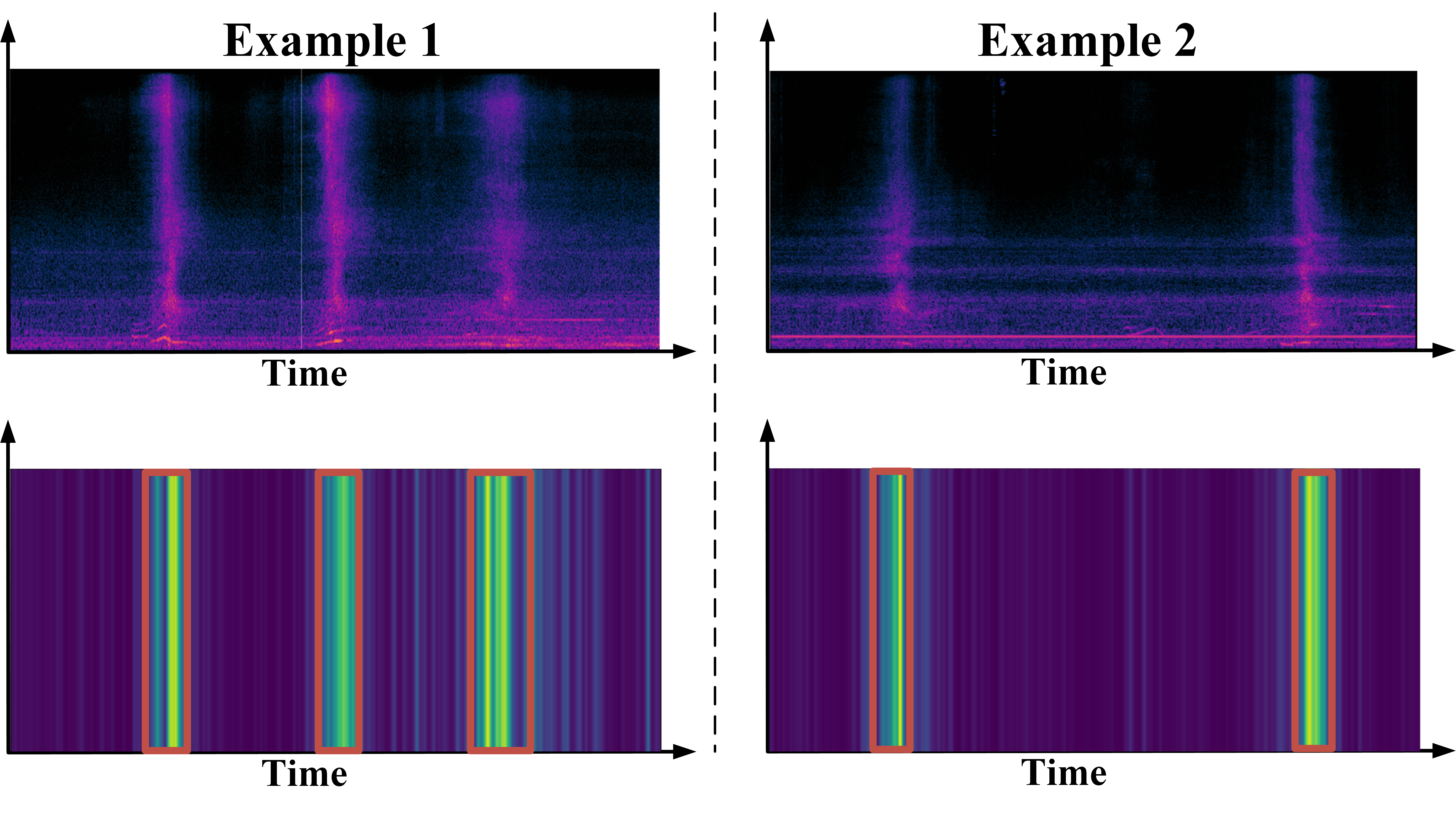}
  \vspace{-4mm}
  \caption{Visualization of graph-enhanced vehicle type feature representation. The top row shows the log-Mel spectrograms of two audio samples, while the bottom row shows the learned corresponding linear interpolation adjacency graphs, with red boxes denoting the attention highlighted vehicle travel events.}
  \label{fig:2}
  \vspace{-5 mm}
\end{figure}

\subsection{Visualization Analysis}
\label{sec:visual}
To demonstrate that our graph-enhanced fined-grained feature representation can capture the contextual association and highlight the important audio events related to vehicle traveling,  we visualize the linear interpolation adjacency graphs of the learned vehicle type feature in Figure~\ref{fig:2}, where we can see the vehicle travel events (i.e., feature nodes) are highlighted in the interpolation adjacency graphs as indicated in the red box areas. The results further validate the effectiveness of our proposed method. 

\section{Conclusion}
\label{sec:page}
In this paper, we have presented a graph-enhanced dual-stream feature fusion network with a pre-trained model for acoustic traffic monitoring, where both vehicle type and direction are taken into account for feature representation. Specifically, a pre-trained model is introduced to mitigate the data scarcity for feature enhancement and graph attention is leveraged for finer type feature representation. Experimental results demonstrate the effectiveness of our proposed method. By fusing fine-grained vehicle type feature and direction feature, our method achieved 1st place in DCASE 2024 Challenge Task 10.

\bibliographystyle{IEEEbib}
\bibliography{strings,refs}
\end{document}